\documentclass[preprint]{elsarticle}

\usepackage{lineno,hyperref}
\usepackage{amsmath}
\usepackage{upgreek}
\usepackage{booktabs}
\usepackage{float}
\usepackage{bm}
\modulolinenumbers[5]

\journal{Nuclear Physics A}

\begin{document}

\begin{frontmatter}

\title{Neutron star equations of state with optical potential constraint}


\author[GSI,TUDA]{S. Anti\'c\corref{cor1}}
\ead{S.Antic@gsi.de}

\author[GSI]{S. Typel}
\ead{S.Typel@gsi.de}

\cortext[cor1]{Corresponding author}

\address[GSI]{GSI Helmholtzzentrum f\"{u}r Schwerionenforschung GmbH,
  \\ Planckstra\ss{}e 1, D-64291 Darmstadt, Germany}
\address[TUDA]{ Technische Universit\"{a}t Darmstadt,
  Schlossgartenstra\ss{}e 2, D-64289 Darmstadt, Germany}




\begin{abstract}
Nuclear matter and neutron stars are studied in the framework of an 
extended relativistic mean-field (RMF) model with
higher-order derivative and density dependent couplings of nucleons 
to the meson fields. The derivative couplings lead to an energy
dependence of the scalar and vector self-energies of the nucleons. It
can be adjusted to be consistent with experimental results for the
optical potential in nuclear matter. Several parametrisations, which give identical
predictions for the saturation properties of nuclear matter,
are presented for different forms of the derivative coupling
functions. The stellar structure of spherical, non-rotating stars 
is calculated for these new equations of state (EoS). A substantial
softening of the EoS and a reduction of the maximum mass of
neutron stars is found if the optical potential constraint is satisfied.
\end{abstract}


\end{frontmatter}


\section{Introduction}

The recent observation of two pulsars with approximately two solar
masses \cite{Demorest:2010bx,Antoniadis:2013pzd} presents a severe
challenge to the theoretical description of cold high-density matter
in $\beta$-equilibrium.
The equation of state (EoS) has to be sufficiently stiff in order to support
such high masses of compact stars. Many models that are solely based on
nucleonic (neutrons and protons) and leptonic (electrons and muons) 
degrees of freedom are able to reproduce maximum neutron star masses
above two solar masses if the effective interaction between the nucleons becomes
strongly repulsive at high baryon densities. However, 
additional hadronic particle species can appear at densities
above two or three times the nuclear saturation density $n_{\rm sat}
\approx 0.16$~fm$^{-3}$. In most cases, these additional degrees of
freedom lead to a substantial softening of the EoS resulting in a reduced
maximum mass of the compact star below the observed values. 
This feature is well-known for
models with hyperons -- the so-called ''hyperon puzzle'', see, e.g., 
\cite{Lonardoni:2014bwa,Fortin:2014mya} and references therein --
but was also observed in approaches 
that take excited states of the nucleons such as $\Delta(1232)$ resonances into
account, see, e.g., \cite{Drago:2014oja,Cai:2015hya} and references therein. 
Usually, only specifically designed interactions can avoid the problem
of too low maximum masses. 

Successful models of the baryonic contribution to the stellar EoS
should be scrutinized whether they comply with other experimental
constraints, e.g.\ with respect to the employed interactions.
In the center of compact stars very high baryon densities are reached 
exceeding several times $n_{\rm sat}$ and the corresponding Fermi
momenta of the particles are much larger than those at saturation.
This is particularly significant for models with only nucleonic degrees of freedom.
Hence, not only the density dependence of the effective in-medium
interaction between nucleons but also their momentum dependence
becomes relevant. For densities near $n_{\rm sat}$ this information is
contained in the optical potential of nucleons that can be extracted
from the systematics of elastic proton scattering on nuclei, see,
e.g., \cite{Hama:1990vr,Cooper:1993nx}. A saturation of the real part
of the optical potential is observed at high kinetic energies
approaching $1$~GeV. The momentum dependence of the in-medium
interaction is also crucial in simulations of heavy-ion collisions 
\cite{Zhang:1994hpa}.

Typical approaches for 
the baryonic contribution to the stellar EoS are
energy density functionals that originate from
nonrelativistic or relativistic mean-field models of nuclear matter.
The most prominent cases among the former class are Skyrme energy
density functionals, see reference \cite{Dutra:2012mb} for an overview of different
parametrizations. They are derived originally from the zero-range Skyrme interaction
with a two-body contribution, which is an expansion up to second order
in the particle momenta, and a density dependent three-body
contribution, which is included in order to reproduce saturation
properties of nuclear matter. Obviously, an extrapolation of the model
to high momenta is questionable given the limited form of the momentum
dependence. Examples of the latter class, frequently denominated covariant
density functionals, can be inferred from relativistic Lagrangian
densities. In conventional models, see reference \cite{Dutra:2014qga}
for a wide collection of different parametrizations, a nucleon optical
potential in the medium can be derived from the relativistic scalar and
vector self-energies, see section \ref{sec:res}. It exhibits a linear increase with
energy, which is in contradiction with the expectation from
experiment. In general, one would expect that the nucleon self-energies 
itself depend explicitly on the particle momentum or energy as, e.g.,
in Dirac-Brueckner calculations of nuclear matter \cite{Fuchs:2003zn}. However, this is
not realized in standard RMF models.

There are particular extensions of relativistic mean-field
(RMF) models that contain
nucleon self-energies with an explicit energy or momentum dependence. 
This dependence cannot be introduced in a relativistic model in a
simple parametric form because it affects, e.g., the definition of the
conserved currents.
In extended systematic approaches new derivative couplings between the nucleon and
meson fields are introduced that allow to reproduce the energy
dependence of the optical potential as extracted from experiments.
One of the earliest RMF models with scalar derivative couplings was
presented in reference \cite{Zimanyi:1990np}. A rescaling of the
nucleon fields removed the explicit momentum dependence of the
self-energies but lead to a considerable softening of the EoS. 
More general couplings of the mesons to linear derivatives of the
nucleon fields were considered in reference \cite{Typel:2002ck} with an
application to uniform nuclear matter. With appropriately chosen
coupling constants a reduction of the optical potential was found 
as compared to the strong linear energy dependence in conventional RMF
models. The model was further extended in reference
\cite{Typel:2005ba} assuming a density dependence of the couplings. It
was successfully applied to the description of finite nuclei. Notable
new features were the increase of the effective nucleon masses 
(usually rather small in order to explain the strong spin-orbit
interaction in nuclei) and correspondingly higher level densities 
close to the Fermi energy in nuclei 
in better concordance with expectations from
experiments. Though, using couplings only linear in the
derivatives leads to a quadratic dependence of the optical potential
on the kinetic energy with a decrease for energies
exceeding $1$~GeV. Couplings to all orders in the derivative of the
nucleons were introduced in the so-called nonlinear derivative (NLD)
model \cite{Gaitanos:2009nt,Gaitanos:2011yb} assuming a particular exponential
dependence on the derivatives but no density dependence of the
nucleon-meson couplings  or self-couplings of the mesons. 
The general formalism was developed and applied to infinite
isospin symmetric and asymmetric nuclear matter with a particular choice of the
non-linear derivative terms that lead to an energy dependence of the self-energies. 
In reference \cite{Gaitanos:2012hg} the approach was extended with 
generalized non-linear derivative couplings of any functional form
in the field-theoretical formalism
allowing for a momentum or energy dependence.
Nonlinear self-couplings of the $\sigma$ meson field were added
in order to improve the
description of characteristic nuclear matter parameters at saturation.
The application of this version of the NLD model to stellar
matter yielded a maximum neutron star mass of $2.03~M_{\rm sol}$ barely satisfying
the observational constraints but the dependence of the result on the
model parameters was not explored in detail. 
The NLD model was also applied to the description of bulk properties of nuclear matter
in reference \cite{Chen:2012rn}. A softening of the EoS and
maximum masses of neutron stars substantially below $2~M_{\rm sol}$
were found with different parametrizations 
assuming an energy dependence of the couplings but 
nonlinear self-couplings of the mesons or density dependent
meson-nucleon couplings were not considered. 
Properties of finite nuclei were studied in reference \cite{Chen:2014xha} 
after adding meson self-interactions in the Lagrangian.
A qualitative description similar to conventional RMF models was achieved but neutron
star properties were not examined in this extended model.

In this work we introduce a more flexible extension of the
nonlinear derivative model assuming density dependent
meson-nucleon couplings in addition. Instead of using derivative operators that generate
an explicit momentum dependence of the self-energies, we will use a functional
form that leads to an energy dependence. This approach will also be more
suitable for a future applications of the DD-NLD approach to nuclei
since the relevant equations and their numerical implementation 
are simplified. Here, the equations of state
of symmetric and asymmetric nuclear matter will be calculated for
different choices of the derivative coupling operators that lead to a
saturation of the optical potential at high energies as derived from experiments. 
They are compared to
the results of a standard RMF model with density dependent couplings
that is consistent with essentially all modern constraints for the 
characteristic nuclear matter parameters at saturation.
The parameters of the DD-NLD models are chosen such that these
saturation properties are reproduced. The effect of the optical
potential constraint on the mass-radius relations of
neutron stars will be studied. 

The paper is organized as follows: In section \ref{sec:model} the
Lagrangian density of the DD-NLD approach is presented. The field
equations in mean-field approximation and the
energy-momentum tensor will be derived. The relevant equations for the
case of infinite nuclear matter will be considered in more detail in
section \ref{sec:inm}. The parametrization of the
density dependent couplings and the functional form of the derivative
coupling functions is discussed in section \ref{sec:para}. Results for
the energy dependence of the optical potential, the EoS of nuclear matter and the
mass-radius relation are presented in section \ref{sec:res} for 
various versions of the model. Conclusions are given in section
\ref{sec:con}. Detailed expression for various densities are collected
in \ref{sec:app_a}.

\section{Lagrangian density and field equations of the DD-NLD model} 
\label{sec:model}

In most RMF models the effective interaction between nucleons is
described by an exchange of mesons. Usually, $\sigma$ and $\omega$
mesons are introduced to consider the attractive and
repulsive contributions to the nucleon-nucleon potential, respectively. They are
represented by isoscalar Lorentz scalar and Lorentz vector fields
$\sigma$ and $\omega_{\mu}$. In order to model the isospin dependence
of the interaction, the exchange of $\rho$ mesons is included. It
is denoted by the isovector Lorentz vector field $\bm{\rho}_{\mu}$ in
the following. The Lagrangian density in the DD-NLD
approach\footnote{Natural units with $\hbar = c = 1$ are used in the following.}
\begin{equation}
 \mathcal{L} = \mathcal{L}_{\rm nuc} + \mathcal{L}_{\rm mes} +
 \mathcal{L}_{\rm int}
\end{equation}
contains contributions of the free nucleons $\Psi = (\Psi_{p},\Psi_{n})$ with mass $m$
\begin{equation}
\label{eq:L_nuc}
 \mathcal{L}_{\rm nuc} = \frac{1}{2} \left( 
 \overline{\Psi}\gamma_{\mu}i\overrightarrow{\partial^{\mu}}\Psi -  
 \overline{\Psi} i \overleftarrow{\partial^{\mu}}\gamma_{\mu}\Psi 
 \right) - m \overline{\Psi}\Psi
\end{equation}
in a symmetrized form and of free mesons
\begin{eqnarray}
 \mathcal{L}_{\rm mes} & = & \frac{1}{2} 
 \left( \partial_{\mu}\sigma\partial^{\mu}\sigma 
 - m^{2}_{\sigma}\sigma^{2}
  - \frac{1}{2}  F^{(\omega)}_{\mu\nu}F^{(\omega)\mu\nu}  
 +  m^{2}_{\omega}\omega_{\mu}\omega^{\mu}
\right. \\ \nonumber & & \left. 
 - \frac{1}{2} \bm{F}^{(\rho)}_{\mu\nu} \bm{F}^{(\rho)\mu\nu} 
 + m^{2}_{\rho} \bm{\rho}_{\mu} \bm{\rho}^{\mu} 
 \right)
\end{eqnarray}
with the field tensors
\begin{equation}
 F^{(\omega)}_{\mu\nu} = \partial_{\mu} \omega_{\nu} - \partial_{\nu}
 \omega_{\mu}
\quad \mbox{and} \quad
 \bm{F}^{(\rho)}_{\mu\nu} = \partial_{\mu} \bm{\rho}_{\nu} - \partial_{\nu}
 \bm{\rho}_{\mu}
\end{equation}
of the isoscalar $\omega$ meson and the isovector $\rho$ meson,
respectively. The arrows in equation (\ref{eq:L_nuc}) denote the
direction of differentiation. 

Standard RMF models assume a minimal coupling of the nucleons to the 
meson fields leading to
\begin{equation}
\label{eq:L_int_st}
\mathcal{L}_{\rm int} =
 \Gamma_{\sigma}  \sigma \overline{\Psi} \Psi  
 - \Gamma_{\omega} \omega_{\mu}\overline{\Psi}\gamma^{\mu}\Psi 
 - \Gamma_{\rho}
 \bm{\rho}_{\mu}\overline{\Psi}\bm{\tau}\gamma^{\mu}\Psi
\end{equation}
for the interaction contribution to the total Lagrangian density
$\mathcal{L}$ with meson-nucleon couplings 
$\Gamma_{i}$ ($i=\sigma,\omega,\rho$). We assume that they depend on
the vector density $n_{v}$, see equation (\ref{eq:n_v}) for the
explicit definition.
In the derivative coupling model the nucleon
field $\Psi$ ($\overline{\Psi}$) is replaced in $\mathcal{L}_{\rm int}$ by $\mathcal{D}_{m}\Psi$ 
($\overline{\mathcal{D}_{m}\Psi}$) with operator functions
$\mathcal{D}_{m}$, which can be different for the various mesons
$m=\sigma,\omega,\rho$. They can be expanded in a series
\begin{equation}
 \mathcal{D}_{m}(x) = \sum_{n=0}^{\infty} \frac{d_{n}^{(m)}}{n!} x^{n} 
\end{equation}
with numerical coefficients $d_{n}^{(m)}$.
The argument $x$ contains derivatives $i\partial_{\beta}$ that
act on the nucleon field. More specifically we write
\begin{equation}
\label{eq:def_x}
 x = v^{\beta}i\partial_{\beta} - sm 
\end{equation}
as a hermitian Lorentz scalar operator 
with an auxiliary Lorentz vector $v^{\beta}=(v_{0},\vec{v})$ and
a scalar factor $s$.
Hence the interaction contribution in the NLD model is written as
\begin{eqnarray}
\mathcal{L}_{\rm int} & = & 
\frac{1}{2} \Gamma_{\sigma} \sigma \left(
  \overline{\Psi}\overleftarrow{\mathcal{D}}_{\sigma}\Psi
 +  \overline{\Psi} \overrightarrow{\mathcal{D}}_{\sigma} \Psi  \right) 
 \\ \nonumber & & 
 - \frac{1}{2} \Gamma_{\omega} \omega_{\mu}\left( 
 \overline{\Psi}\overleftarrow{\mathcal{D}}_{\omega}\gamma^{\mu}\Psi 
+ \overline{\Psi}\gamma^{\mu}\overrightarrow{\mathcal{D}}_{\omega}\Psi \right)
  \\ \nonumber & & 
 -  \frac{1}{2} \Gamma_{\rho} \bm{\rho}_{\mu} \left(
   \overline{\Psi}\overleftarrow{\mathcal{D}}_{\rho}\gamma^{\mu}\bm{\tau}\Psi
 + \overline{\Psi}\bm{\tau}\gamma^{\mu}\overrightarrow{\mathcal{D}}_{\rho}\Psi
\right)
\end{eqnarray}
in a symmetrized form with respect to the derivative operators $\mathcal{D}_{m}$,
i.e.,
\begin{eqnarray}
 \overrightarrow{\mathcal{D}}_{m} &=& \sum^{\infty}_{k=0} C_{k}^{(m)}
 (v^{\beta }i \overrightarrow{\partial}_{\beta})^{k} \\
 \overleftarrow{\mathcal{D}}_{m} &=& \sum^{\infty}_{k=0} C_{k}^{(m)}
 (-v^{\beta }i \overleftarrow{\partial}_{\beta})^{k} 
\end{eqnarray}
with coefficients
\begin{equation}
 C_{k}^{(m)} = \sum^{k}_{n=0}  \frac{d_{n}^{(m)}}{n!} \binom{n}{k} (-sm)^{n-k} \: .
\end{equation}
Obviously, no derivatives appear for the choice $\mathcal{D}_{m}=1$ (corresponding to
$d_{n}^{(m)} = \delta_{n0}$) and the standard form (\ref{eq:L_int_st}) is
recovered.

When spatially inhomogeneous systems with Coulomb interaction are
considered, $\mathcal{L}_{\rm mes}$ and $\mathcal{L}_{\rm int}$ can be
complemented with the appropriate contributions. Since only
uniform matter is considered in the following, we do not give them
here explicitly.

The field equations of nucleons are derived from the
generalized Euler-Lagrange equation
\begin{equation}
\label{eq:EuLa}
\frac{\partial\mathcal{L}}{\partial \varphi_r} 
+ \sum^{\infty}_{i=1}(-)^i \partial_{\alpha_1,\dots,\alpha_i}
 \frac{\partial\mathcal{L}}{\partial(\partial_{\alpha_1,\dots,\alpha_i}\varphi_r)} = 0
\end{equation}
for $\varphi_{r}=\Psi, \overline{\Psi}$. For the meson fields $\varphi_{r}= \sigma, \omega_{\mu},
\bm{\rho}_{\mu}$ the standard Euler-Lagrange equation applies, i.e.\
only the $i=1$ term in equation (\ref{eq:EuLa}) is relevant since higher-order 
derivatives of the meson fields do not appear in the Lagrangian density $\mathcal{L}$.

Details can found in references
\cite{Gaitanos:2009nt,Gaitanos:2012hg}.
The Dirac equation
\begin{equation}
\label{eq:Dirac}
 \left[\gamma_{\mu} \left(i\partial^{\mu} - \Sigma^{\mu}\right) 
 - \left( m - \Sigma\right)\right] \Psi = 0
\end{equation}
for the nucleons looks formally the same as in standard RMF approaches
but the scalar ($\Sigma$) and vector ($\Sigma^{\mu}$) self-energy operators now
contain the derivative operators $\mathcal{D}_{m}$. They are given by
\begin{equation}\
\label{eq:Sigma_s}
\Sigma = \Gamma_{\sigma} \sigma \overrightarrow{\mathcal{D}}_{\sigma} 
\end{equation}
and
\begin{equation}
\label{eq:Sigma_v}
\Sigma^{\mu}= \Gamma_{\omega}\omega^{\mu}\overrightarrow{\mathcal{D}}_{\omega} 
+ \Gamma_{\rho}\bm{\tau} \cdot \bm{\rho}^{\mu} 
\overrightarrow{\mathcal{D}}_{\rho} + \Sigma^{\mu}_{R} 
\end{equation}
with the 'rearrangement' contribution
\begin{eqnarray}
\Sigma^{\mu}_{R} & = &
\frac{j^{\mu}}{n_{v}} \left[ 
 \Gamma_{\omega}^{\prime} 
 \omega^{\nu} 
 \frac{1}{2} \left( \overline{\Psi}  \overleftarrow{\mathcal{D}}_{\omega} \gamma_{\nu} \Psi
 + \overline{\Psi} \gamma_{\nu} \overrightarrow{\mathcal{D}}_{\omega} \Psi \right)
 \right. \\ \nonumber & & \left.
 + \Gamma_{\rho}^{\prime} 
 \bm{\rho}^{\nu}  \frac{1}{2} \left(
  \overline{\Psi} \overleftarrow{\mathcal{D}}_{\rho}  \gamma_{\nu} \bm{\tau} \Psi
 + \overline{\Psi} \gamma_{\nu} \bm{\tau} \overrightarrow{\mathcal{D}}_{\rho}
 \Psi \right)
 \right. \\ \nonumber & & \left.
 - \Gamma_{\sigma}^{\prime} \sigma  \frac{1}{2}  
 \left(   \overline{\Psi} \overleftarrow{\mathcal{D}}_{\sigma} \Psi 
 + \overline{\Psi} \overrightarrow{\mathcal{D}}_{\sigma} \Psi  \right) \right] 
\end{eqnarray}
containing derivatives
\begin{equation}
 \Gamma_{i}^{\prime} =  \frac{d\Gamma_{i}}{dn_{v}}
\end{equation}
of the coupling functions.
In the case of inhomogeneous systems and a non-vanishing three-vector
component $\vec{v}$ of the auxiliary vector $v^{\beta}$, additional
contributions in (\ref{eq:Sigma_s}) and (\ref{eq:Sigma_v}) will
appear. In the present application of the DD-NLD model, however, we
do not consider this case.
The field equations of the mesons are found as
\begin{eqnarray}
 \partial_{\mu}\partial^{\mu} \sigma + m^{2}_{\sigma} \sigma 
 & = & \frac{1}{2} \Gamma_{\sigma} \left( 
 \overline{\Psi} \overleftarrow{\mathcal{D}}_{\sigma} \Psi 
+ \overline{\Psi}\overrightarrow{\mathcal{D}}_{\sigma} \Psi  \right) 
 \\ 
\partial_{\mu} F^{(\omega)\mu\nu} + m^{2}_{\omega} \omega^{\nu} 
 & = & \frac{1}{2} \Gamma_{\omega} \left( 
 \overline{\Psi} \overleftarrow{\mathcal{D}}_{\omega} \gamma^{\nu} \Psi 
+ \overline{\Psi}\gamma^{\nu} \overrightarrow{\mathcal{D}}_{\omega} \Psi \right)
 \\ 
 \partial_{\mu} \bm{F}^{(\rho)\mu\nu} + m^{2}_{\rho} \bm{\rho}^{\nu} 
 & = & \frac{1}{2} \Gamma_{\rho} \left(
 \overline{\Psi} \overleftarrow{\mathcal{D}}_{\rho} \gamma^{\nu} \bm{\tau}\Psi 
+ \overline{\Psi}\bm{\tau}\gamma^{\nu}
\overrightarrow{\mathcal{D}}_{\rho} \Psi \right)
\end{eqnarray}
with source terms containing derivative operators.

The conserved baryon current in the DD-NLD model is given by
\begin{equation}
\label{eq:J}
 J^{\mu} = \sum_{i=p,n} \langle\overline{\Psi}_{i}N^{\mu}\Psi_{i}\rangle
\end{equation}
with the norm operator
\begin{equation}
\label{eq:norm}
 N^{\mu}  =  \gamma^{\mu} + \Gamma_{\sigma} \sigma 
  \left(\partial^{\mu}_{p} \mathcal{D}_{\sigma} \right)  
 -  \Gamma_{\omega}\omega_{\alpha} \gamma^{\alpha}
 \left( \partial^{\mu}_{p} \mathcal{D}_{\omega}\right)  
 -  \Gamma_{\rho} \bm{\rho} _{\alpha} 
  \gamma^{\alpha} \bm{\tau} \left( \partial^{\mu}_{p}
    \mathcal{D}_{\rho}\right)
\end{equation}
where $\partial^{\mu}_{p}\mathcal{D}_{m}$ is the derivative of $\mathcal{D}_{m}$
operator with respect to the momentum $p_{\mu}=i\partial_{\mu}$, i.e.\
\begin{equation}
  \partial^{\mu}_{p}\mathcal{D}_{m} 
 = v^{\mu}\sum^{\infty}_{k=1} kC_{k}^{(m)} (v^{\beta }i\partial_{\beta})^{k-1} 
 \: ,
\end{equation}
and $\langle \dots \rangle$ denotes the summation over all occupied
states.
The current (\ref{eq:J}) is not identical to the vector current
\begin{equation}
\label{eq:J_v}
 J_{v}^{\mu} = \sum_{i=p,n}
 \langle\overline{\Psi}_{i}\gamma^{\mu}\Psi_{i}\rangle \: ,
\end{equation}
which is used to define the vector density
\begin{equation}
\label{eq:n_v}
 n_{v} = \sqrt{J_{v}^{\mu}J_{v\mu}}
\end{equation}
appearing as the argument of the coupling functions $\Gamma_{i}$.
The energy-momentum tensor assumes the form
\begin{equation}
 T^{\mu \nu}  =  \sum_{i=p,n} 
   \langle\overline{\Psi}_{i}N^{\mu} p^{\nu} \Psi_{i}\rangle 
- g^{\mu\nu}\langle \mathcal{L} \rangle \: .
\end{equation}
Then the energy density $\varepsilon$ and pressure $p$ are found from
$\varepsilon = T^{00}$ and $p = \sum_{i=1}^{3}T^{ii}/3$, respectively.

\section{DD-NLD model for nuclear matter}
\label{sec:inm}

In the case of stationary nuclear matter, the equations simplify considerably
since the system is homogeneous and the meson fields, which are
treated as classical fields, are constant in space and
time. Positive-energy solutions of the
Dirac equation (\ref{eq:Dirac}) are plane waves $\Psi_{i} = u_{i}
\exp \left( - i p_{i}^{\mu} x_{\mu}\right)$ for protons
and neutrons with Dirac spinors $u_{i}$, which are normalized according to
\begin{equation}
 \overline{\Psi}_{i} N^{0}
 \Psi_{i} = \bar{u}_{i} N^{0} u_{i}  = 1
\end{equation}
with the time component of the norm operator (\ref{eq:norm}).
They depend on the effective mass
\begin{equation}
 m_{i}^{\ast} = m_{i} - \Sigma_{i}
\end{equation}
and effective momentum
\begin{equation}
 p_{i}^{\ast\mu} = p_{i}^{\mu} - \Sigma_{i}^{\mu} 
\end{equation}
related by the dispersion relation
\begin{equation}
 p_{i}^{\ast\mu} p_{i\mu}^{\ast} = \left(m_{i}^{\ast}\right)^{2} \: .
\end{equation}
The derivative $i\partial^{\beta}$ in the $\mathcal{D}_{m}$ operators
can be replaced by the corresponding four-momentum
$p_{i}^{\beta}=(E_{i},\vec{p}_{i})$
resulting in a simple function $D_{m}$ depending on the energy $E_{i}$ and
the momentum $\vec{p}_{i}$ of the nucleon. 

Using the identity
\begin{equation}
 N^{\mu} \Psi_{i} = \left[ \gamma^{\mu} + \partial_{p}^{\mu}
   \Sigma_{i} - \gamma_{\alpha} \partial_{p}^{\mu}  \left(
     \Sigma_{i}^{\alpha} - \Sigma_{R}^{\alpha} \right)
 \right] \Psi_{i}
\end{equation}
the conserved current and the energy-momentum tensor can be written as
\begin{equation}
 J^{\mu} = \sum_{i=p,n} \kappa_{i}
\int \frac{d^{3} p}{(2\pi)^{3}} \: \frac{\Pi^{\mu}_{i}}{\Pi^{0}_{i}}
\end{equation}
and
\begin{equation}
 T^{\mu\nu}= \sum_{i=p,n} \kappa_{i} \int \frac{d^{3}p}{(2\pi)^{3}} \: 
 \frac{\Pi^{\mu}_{i} p^{\nu}}{\Pi^{0}_{i}} -
 g^{\mu\nu}\langle\mathcal{L}\rangle \: ,
\end{equation}
respectively, with the four-momentum
\begin{equation}
\Pi^{\mu}_{i} = p^{\ast\mu}_{i} + m^{\ast}_{i} \left(\partial^{\mu}_{p}\Sigma_{i}\right)
  - p^{\ast}_{i\beta} \left[\partial^{\mu}_{p} \left( \Sigma^{\beta}_{i}-
      \Sigma^{\beta}_{R}\right)\right]
\end{equation}
and spin degeneracy factors $\kappa_{i}=2$.
The integration runs over all momenta $p$ with modulus lower than the
Fermi momenta $p_{Fi}$ in the no-sea approximation. They are defined through the individual
nucleon densities
\begin{equation}
\label{eq:n_i}
 n_{i} = \frac{\kappa_{i}}{6\pi^{2}} p_{Fi}^{3} \: .
\end{equation}

Without the preference for a particular direction in infinite nuclear
matter, the spatial components of
the Lorentz vector meson fields vanish and the auxiliary vector in
equation (\ref{eq:def_x}) is set to $v^{\beta}=\delta_{\beta 0}$ such
that the $D_{m}$ functions only depend on the nucleon energy $E_{i}$.
Without isospin changing processes, only the third
component of the isovector $\rho$ field has to be considered in the
field equations for the mesons. Using the abbreviations
$\omega = \omega^{0}$ and $\rho = \bm{\rho}^{0}_{3}$ the meson fields
are immediately obtained from
\begin{eqnarray}
\label{eq:feq_sigma}
\sigma & = & \frac{\Gamma_{\sigma}}{m_{\sigma}^{2}} n_{\sigma}
 =  \frac{\Gamma_{\sigma}}{m_{\sigma}^{2}} \sum_{i=p,n} \langle \overline{\Psi}_{i} D_{\sigma}
 \Psi_{i} \rangle
 \\
\label{eq:feq_omega}
\omega & = & \frac{\Gamma_{\omega}}{m_{\omega}^{2}} n_{\omega} 
 = \frac{\Gamma_{\omega}}{m_{\omega}^{2}} \sum_{i=p,n} \langle \overline{\Psi}_{i} \gamma^{0} D_{\omega}
 \Psi_{i} \rangle
 \\
\label{eq:feq_rho}
\rho & = & \frac{\Gamma_{\rho}}{m_{\rho}^{2}} n_{\rho} 
 = \frac{\Gamma_{\rho}}{m_{\rho}^{2}}  \sum_{i=p,n} \langle \overline{\Psi}_{i} \gamma^{0}
 \tau_{3} D_{\rho}  \Psi_{i} \rangle
\end{eqnarray}
with source densities $n_{\sigma}$, $n_{\omega}$, and $n_{\rho}$.
The self-energies simplify to
\begin{eqnarray}
 \Sigma_{i} & = & \Gamma_{\sigma} \sigma D_{\sigma}
\\ 
 \Sigma_{i}^{0} & = & \Gamma_{\omega}\omega D_{\omega}
 + \Gamma_{\rho} \tau_{3,i} \rho D_{\rho}
 + \Sigma^{0}_{R} 
\\
 \vec{\Sigma}_{i} & = & 0
\end{eqnarray}
with $\tau_{3,i} = 1$ ($-1$) for protons (neutrons) and
the 'rearrangement' contribution
\begin{equation}
\Sigma^{0}_{R} = 
 \Gamma_{\omega}^{\prime}  \omega n_{\omega} 
 + \Gamma_{\rho}^{\prime} \rho n_{\rho}
 - \Gamma_{\sigma}^{\prime} \sigma n_{\sigma}
\end{equation}
is independent of the nucleon energy.
The dispersion relation reads 
\begin{equation}
\label{eq:disp}
 E_{i} = \sqrt{p^{2}+\left( m_{i} - S_{i} \right)^{2}} + V_{i} 
\end{equation}
if we introduce the energy-dependent scalar potentials $S_{i}(E) = \Sigma_{i}$ and
vector potentials $V_{i}(E) = \Sigma_{i}^{0}$.
Explicit expressions for the various densities and
thermodynamic quantities of the DD-NLD model are given in \ref{sec:app_a}.

\section{Parametrization of the DD-NLD model} 
\label{sec:para}

For the application of the DD-NLD model to nuclear matter the parameters need to be
specified. Besides the usual parameters of a RMF model with density
dependent couplings the form of the $D_{m}$ functions has to be given.
We assume identical functions for all mesons, i.e.\ $D = D_{\sigma} =
D_{\omega} = D_{\rho}$, and consider three functional dependencies:
\begin{itemize}
\item[D1] a constant $D = 1$, which corresponds to a usual RMF model with density
  dependent couplings,
\item[D2] a Lorentzian form $D = 1/(1+x^{2})$,
\item[D3] an exponential dependence $D=\exp\left( -x \right)$
\end{itemize}
with $x=(E_{i}-m_{i})/\Lambda$ because we set
$v^{\beta}=\delta_{\beta  0}$ and $s=1$ in equation
(\ref{eq:def_x}). The parameter $\Lambda$ regulates the strength of
the energy dependence.
For the proton and neutron masses the experimental values
of $m_{p} = 938.272046$~MeV/c${}^{2}$ and $m_{n} = 939.565379$~MeV/c${}^{2}$,
respectively, are used. The meson masses are set to
$m_{\sigma}=550$~MeV/c${}^{2}$, $m_{\omega}=783$~MeV/c${}^{2}$, and $m_{\rho}=763$~MeV/c${}^{2}$. 
The density dependence of the meson nucleon couplings has the same
form as introduced in reference \cite{Typel:1999yq}. 
We assume a dependence on the vector density $n_{v}$ as defined
in equation (\ref{eq:n_v}), which is not identical to the zero-component of the
conserved baryon current (\ref{eq:J}). See \ref{sec:app_a} for explicit expressions.
This choice simplifies the rearrangement terms considerably.   

For the isoscalar
mesons $m=\sigma,\omega$ the coupling is written as
\begin{equation}
 \Gamma_{m}(n_{v}) = \Gamma_{m}(n_{\rm ref}) f_{m}(x) 
\end{equation}
with functions
\begin{equation}
 f_{m}(x) = a_{m} \frac{1 + b_{m} (x + d_{m})^{2}}{1 + c_{m} (x +d_{m})^{2}}
\end{equation}
that depend on the argument $x=n_{v}/n_{\rm ref}$ and contain
coefficients $a_{m}$, $b_{m}$, $c_{m}$, and $d_{m}$. For the 
$\rho$-meson coupling we set
\begin{equation}
 \Gamma_{\rho}(n_v) = \Gamma_{\rho}(n_{\rm ref}) \exp\left[-a_{\rho}
   \left(x-1\right)\right] \: .
\end{equation}
In order to reduce the number of independent parameters we demand that
the conditions $f_{\sigma}(1)=f_{\omega}(1)=1$ and $f_{\sigma}^{\prime\prime}(0) =
f_{\omega}^{\prime\prime}(0)=0$ hold. Hence, there are only two
independent coefficients in the functions $f_{m}$ for each of the isoscalar mesons.
The overall magnitude of the couplings is given by
the couplings $\Gamma_{m}(n_{\rm ref})$ at a reference density
$n_{\rm ref}$. 
We require that the characteristic saturation properties for the three
choices of the $D$ function are identical and close to current values
extracted from experiments. 
In particular, we set the saturation density to $n_{\rm sat}=
0.15$~fm${}^{-3}$, the binding energy per nucleon at saturation to
$B=16$~MeV, the compressibility to $K = 240$~MeV, the symmetry energy
to $J = 32$~MeV and the symmetry energy slope coefficient to $L=60$~MeV.
Furthermore, we set the effective nucleon mass at saturation to
$m_{\rm eff} = 0.5625~m_{\rm nuc}$ (related to the strength of the
spin-orbit potential in nuclei) and fix the ratios 
$f_{\omega}^{\prime}(1)/f_{\omega}(1)=-0.15$ and
$f_{\omega}^{\prime\prime}(1)/f_{\omega}^{\prime}(1)=-1.0$ in order
to determine the coefficients in the functions $f_{m}$ uniquely. 
These values are close to those of the parametrization DD2 \cite{Typel:2009sy}
that was fitted to properties of nuclei and predicts a neutron star maximum mass of
$2.4~M_{\rm sol}$. 

\begin{table}[t]
 \caption{\label{tab:1}%
Parameters of the meson coupling functions for three choices of the $D$
functions and different values of cut-off parameter 
 $\Lambda$.}
 \centering
 \begin{tabular}{clccccc}
 \toprule 
  meson & parameter  & \multicolumn{1}{c}{D1}
  &\multicolumn{2}{c}{D2} & \multicolumn{2}{c}{D3}
 \\ 
 \midrule
 &  $\Lambda$ [MeV] & $-$ &  $400$ & $500$ &  $600$ & $700$ \cr
 \midrule
 $\sigma$ & $\Gamma_{\sigma}(n_{\rm ref})$  & 10.72913 & 10.93466 & 10.86315 &  9.74679& 9.89158 \cr
 & $a_{\sigma}$ &  1.36402 &  1.35816 & 1.36015 &  1.38410 & 1.38064 \cr
 & $b_{\sigma}$ &  0.53404 & 0.51914 & 0.52433 &  0.61515 & 0.60127 \cr
 & $c_{\sigma}$ & 0.86714 &  0.83989 & 0.84931 &  1.00615 & 0.98211 \cr
 & $d_{\sigma}$ & 0.62000 &  0.62998 & 0.62648 &  0.57558 &  0.58258 \cr
\midrule
  $\omega$ & $\Gamma_{\omega}(n_{\rm ref})$  &13.29858 & 13.56462 & 13.47215 & 12.0503 & 12.23457 \cr
 & $a_{\omega}$ &  1.3822 &  1.3822 & 1.3822 &  1.3822 & 1.3822 \cr
 & $b_{\omega}$ &  0.42253 &  0.42253 & 0.42253 &  0.42253 & 0.42253 \cr
 & $c_{\omega}$ &  0.71932 &  0.71932 & 0.71932 & 0.71932 & 0.71932 \cr
 & $d_{\omega}$ &  0.60473 &  0.68073 & 0.68073 &  0.68073 & 0.68073 \cr
\midrule
$\rho$ & $\Gamma_{\rho}(n_{\rm ref})$ &  3.59367 &  3.67852& 3.64957 & 3.18819 & 3.25233 \cr
 & $a_{\rho}$ &  0.48762 &  0.48954 & 0.48872 &  0.34777 & 0.36279\cr
 \midrule
 & $n_{\rm ref}$ [fm$^{-3}$] & 0.15000 & 0.14618 & 0.147485 &  0.16515 & 0.16268\cr
 \bottomrule
\end{tabular}
\end{table}
 
Explicit values of the model parameters are given
in table \ref{tab:1} with two choices of the cut-off parameter
$\Lambda$ for the cases of Lorentzian and exponential functions $D$. 
Note that the coefficients of the function $f_{\omega}$ are identical for all five
parametrizations due to the constraints. The reference density
$n_{\rm ref}$ is not necessarily identical to the saturation density
$n_{\rm sat}$ because in the case of explicit derivative couplings the
vector density $n_{v}$ is different from the conserved baryon density
$n_{B} = J^{0}=n_{p}+n_{n}$.

\section{Results}
\label{sec:res}

\begin{figure}[t]
  \centering
  \includegraphics[width=0.80\textwidth]{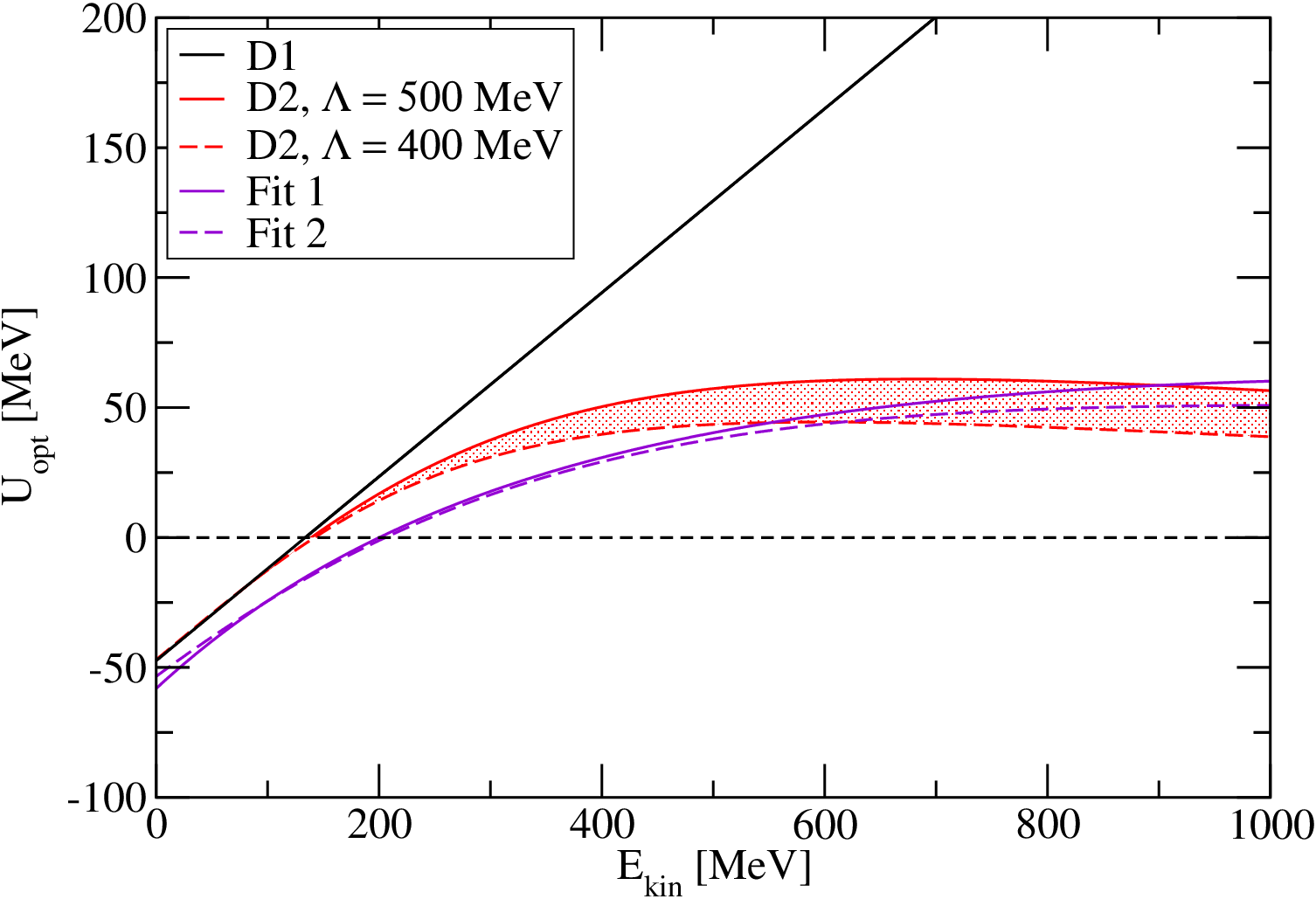}
  \caption{\label{fig:1}%
    The optical potential $U_{\rm opt}$ as a function of the kinetic
    energy $E_{\rm kin}=E-m_{\rm nuc}$ of a nucleon in symmetric nuclear matter at
    saturation density in RMF models with parametrizations D1 and
    D2 compared to two fits from Dirac phenomenology. See text for
    details.}
\end{figure}

\begin{figure}[t]
  \centering
  \includegraphics[width=0.80\textwidth]{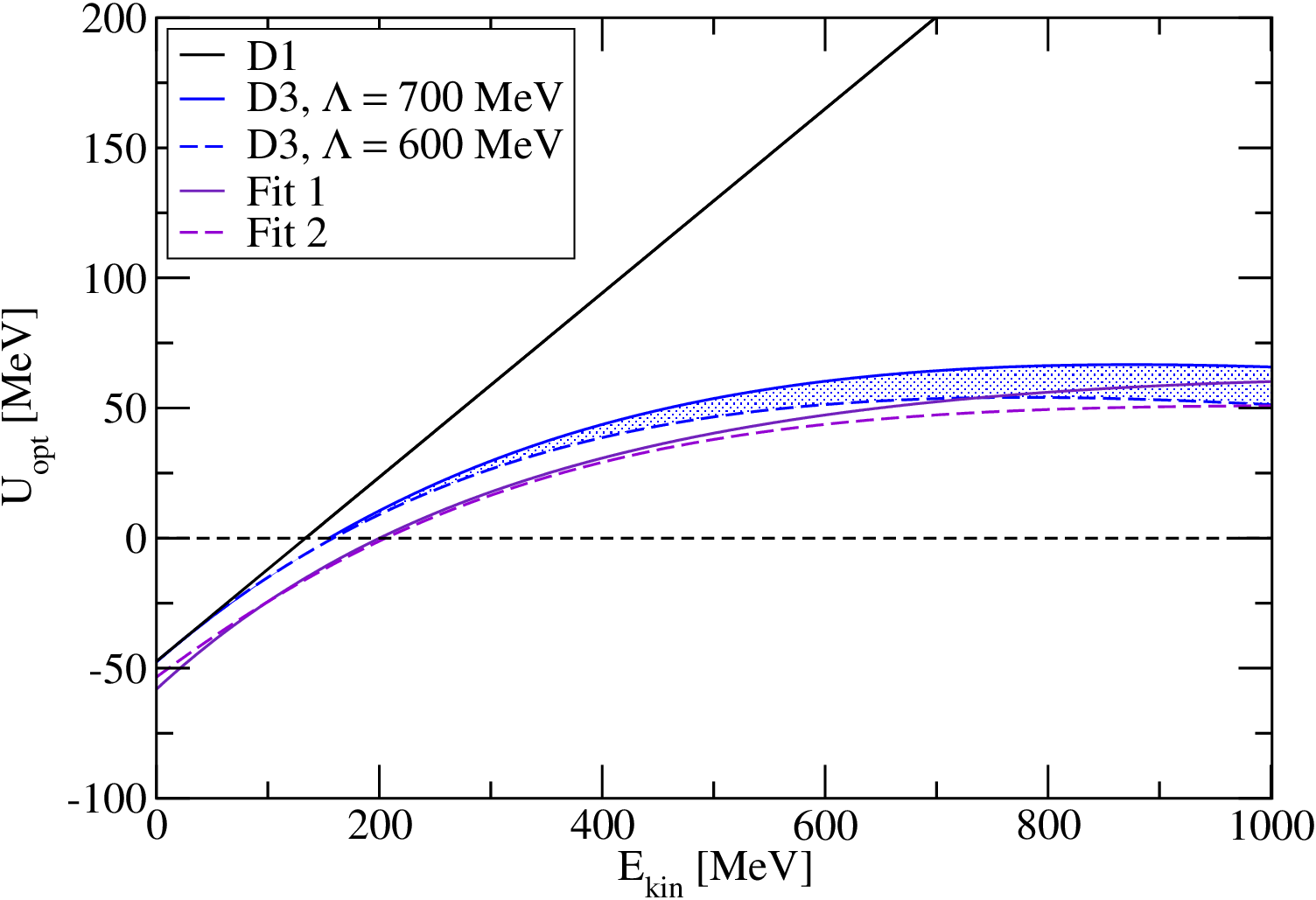}
  \caption{\label{fig:2}%
    The optical potential $U_{\rm opt}$ as a function of the kinetic
    energy $E_{\rm kin}$ of a nucleon in symmetric nuclear matter at
    saturation density in RMF models with parametrizations D1 and
    D3 compared to two fits from Dirac phenomenology. See text for
    details.}
\end{figure}

The nonlinear derivative couplings are introduced in the
RMF model in order to improve the energy dependence of
the optical potential $U_{\rm opt}$. The elastic proton
scattering on nuclei of different mass number $A$ can be well described
in Dirac phenomenology with scalar ($S$) and vector ($V$) potentials, which smoothly
vary with $A$ and the energy of the projectile
\cite{Hama:1990vr,Cooper:1993nx}. From these global fits
the optical potential in symmetric nuclear
matter at saturation density is
obtained as a function of the kinetic energy $E_{\rm kin} = E - m_{\rm
nuc}$ in the limit $A\to \infty$. There are different definitions of the
nonrelativistic optical potential when it is derived from relativistic
scalar and vector self-energies. Here we use the form
\begin{equation}
\label{eq:U_opt}
 U_{\rm opt}(E) = \frac{E}{m_{\rm nuc}}V -S + \frac{S^{2} - V^{2}}{2m_{\rm
     nuc}} 
\end{equation}
with $S=\Sigma_{p}$ and $V = \Sigma_{p}^{0}$ as in references
\cite{Typel:2002ck,Typel:2005ba,Gaitanos:2009nt,Gaitanos:2011yb,Gaitanos:2012hg}.
In conventional RMF models without derivative couplings, the scalar
and vector potentials are constant in energy and the optical potential
(\ref{eq:U_opt}) is just a linear function in energy. This is clearly
seen in figures \ref{fig:1} and \ref{fig:2} as a full black line for
the calculation with the parametrization D1. In contrast, the optical potentials
derived from the scalar and vector potentials in Dirac phenomenology
from two different fits \cite{Hama:1990vr} are much smaller at high energies and exhibit a
saturation for $E_{\rm kin}$ approaching $1$~GeV. At low energies, the optical
potentials from experiment behave more similar as that of the 
theoretical model concerning the absolute strength and
the energy dependence. In figure \ref{fig:1} (\ref{fig:2}) the result for
the DD-NLD parametrization D2 (D3) is depicted for two values of the
cut-off parameter $\Lambda$. Here, a reasonable description of the
experimental optical potential is achieved due to the energy
dependence of the nucleon self-energies. The difference between
parametrizations D2 and D3 is not very significant. The dependence of
$U_{\rm opt}$ on $\Lambda$ is stronger for D2. The deflection of the
DD-NLD curve for that of the standard RMF model D1 appears at lower
kinetic energies for the D3 parametrization as compared to the D2
case. In the DD-NLD model, the optical potential can be calculated
easily for other baryon densities and arbitrary neutron-proton
asymmetries. Because there are no experimental data available for
these general cases, we refrain from presenting the results here. 
But see reference \cite{Typel:2002ck} for the systematics with density in
the linear derivative coupling model.

\begin{figure}[t]
 \centering
 \includegraphics[width=0.8\textwidth]{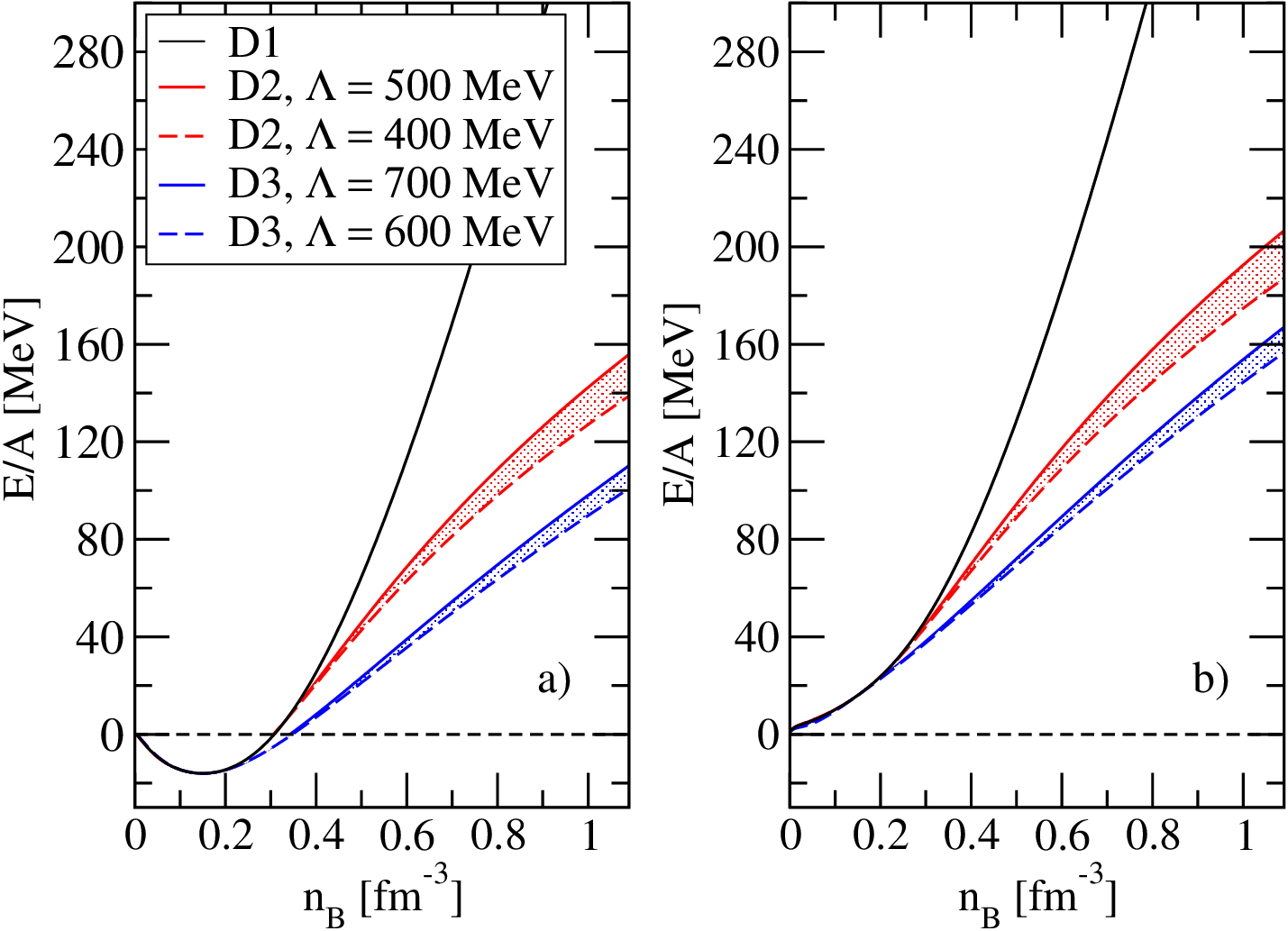}
 \caption{\label{fig:3}%
   Energy per nucleon $E/A$ as a function of the baryon
   density $n_{B}$ for symmetric nuclear matter (a) and pure neutron matter (b).
  Results are given for three different choices of the $D$ function
  and different values of the cut-off parameter $\Lambda$.}
\end{figure}

The reduction of the optical potential at high kinetic energies, which
originates from the energy dependence of the self-energies, is also
reflected in the equation of state. In figure \ref{fig:3} the 
energy per nucleon $E/A$ (without the rest mass contribution) 
is depicted as a function of the baryon
density $n_{B}$ in symmetric nuclear matter (left panel) and neutron
matter (right panel). In both cases, a substantial softening of the
EoS is found as compared to the standard RMF calculation with
parametrisation D1. The effect is stronger for an exponential energy
dependence of the self-energies (D3) than for the case of a Lorentzian
dependence (D2). By construction, all EoS are identical at the
saturation density $n_{\rm sat}$.

\begin{figure}[t]
   \centering
   \includegraphics[width=0.8\textwidth]{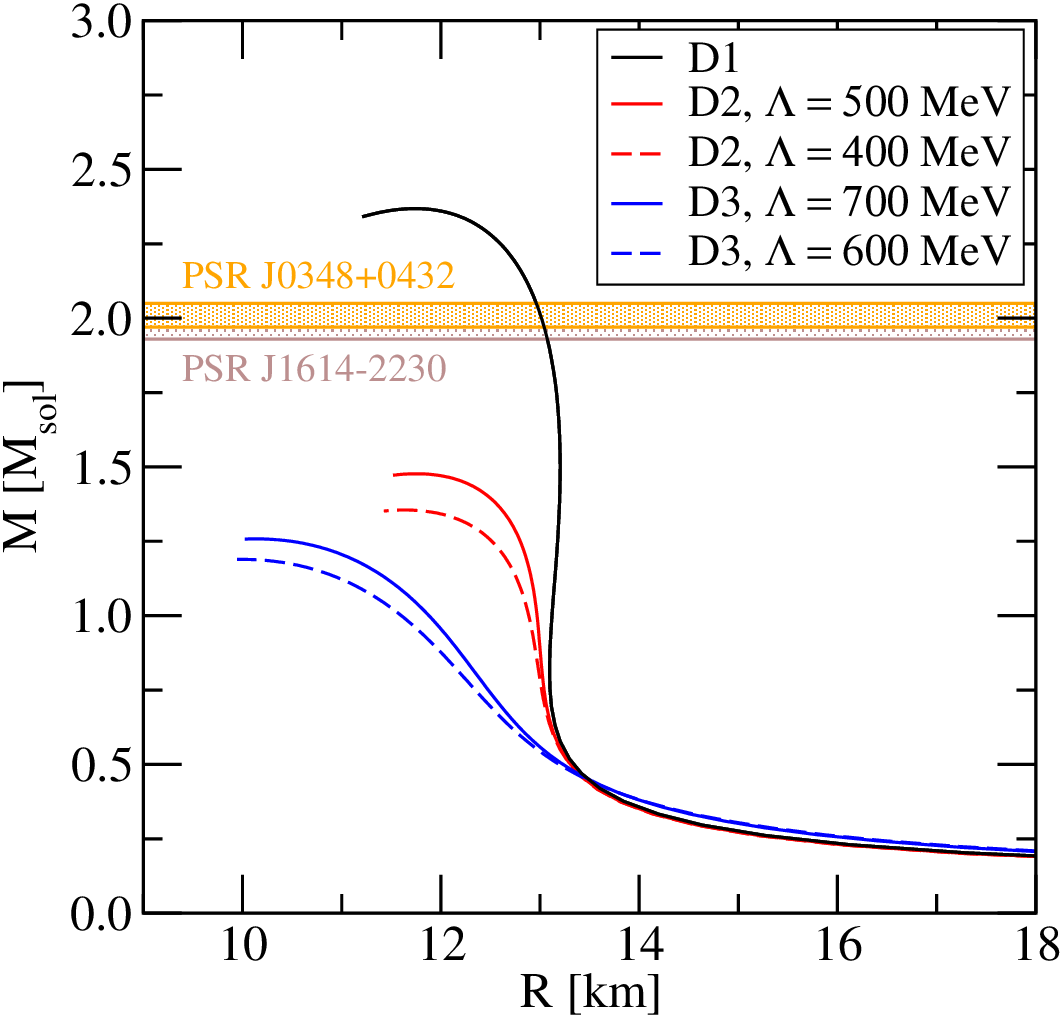}
   \caption{\label{fig:4}%
   Mass-radius relation of neutron stars for different
     choices of the $D$ functions and cut-off parameters $\Lambda$ in the
     DD-NLD model. The two shaded bands refer to astrophysical mass measurements of the pulsars
    PSR~$J1614-2230$ \cite{Demorest:2010bx} and PSR~$J0348+0432$ \cite{Antoniadis:2013pzd}.}
   \label{MR}
\end{figure}

The DD-NLD model can be used to predict the properties of
neutron stars. Here, the EoS of stellar matter is required. It is
obtained by adding the contribution of electrons to the 
energy density and pressure of the baryons. The conditions of charge
neutrality and $\beta$ equilibrium fix the lepton density and
proton-neutron asymmetry uniquely. Since the present model calculations treat
only homogeneous matter, a suitable EoS for the crust of neutron stars
has to be added at low densities. We use the standard
Baym-Pethick-Sutherland (BPS) crust EoS \cite{Baym:1971pw}.
The mass-radius relation of
neutron stars is found finally by solving the
Tolman-Oppenheimer-Volkoff equations
\cite{Oppenheimer:1939ne,Tolman:1939jz}.
It is shown for the five models of this work in figure \ref{fig:4}
together with the masses of the two most massive pulsars observed so
far. The model without an energy dependence (black full line) can explain without any
difficulties the large neutron star masses from astrophysical
observations \cite{Demorest:2010bx,Antoniadis:2013pzd} because the EoS
is rather stiff at high densities. In contrast, for the EoS of neutron
star matter in the DD-NLD models that are consistent with the optical
potential constraint, a serious reduction of the maximum neutron star
mass is seen. 
These models have even problems to reach typical masses of about
$1.4~M_{\rm sol}$ of ordinary neutron stars. Deviations from the
predictions of the standard model D1 start to appear already at masses
below $0.7~M_{\rm sol}$. There is also an effect on the neutron star
radius, which is found to be smaller in the D2 and in
the D3 model. Explicit values for the maximum mass as well as the
radius and central density at this extreme conditions are given in
table \ref{tab:2}. For the DD-NLD models D2 and, in particular, D3 the
central densities is a star of maximum mass are considerably higher
than those of the standard RMF model without an energy dependence of the couplings.

It is worthwhile to compare our results for the mass-radius relation of neutron stars
with those of previous versions of the NLD model. In the approach of 
reference \cite{Gaitanos:2012hg}
it was possible to reach a maximum neutron star mass of about $2~M_{\rm sol}$ with their choice
of parameters. In contrast, the results of reference \cite{Chen:2012rn} using different
functional forms of the couplings indicate a reduction of the maximum mass below the observed
values in line with our calculations. All three versions of the NLD model are adjusted
to similar values of the nuclear matter parameters at saturation, such as saturation density,
binding energy, compressibility or symmetry energy consistent with experimental constraints. 
Nevertheless, the predictions for matter properties at supra-saturation densities are rather
different due to the various choices in the models to represent the effective 
in-medium interaction.
In our paper, an explicit density dependence of the couplings is considered whereas in
\cite{Gaitanos:2012hg} nonlinear self-couplings of the meson fields were
assumened. In the approach of \cite{Chen:2012rn} to nuclear matter neither meson self-couplings
nor a density dependence of the couplings were used.
A reasonable description of nuclear matter near saturation does not 
determine the high-density behaviour of the EoS uniquely, in particular due to the
additional freedom with the 
explicit momentum/energy dependence in the NLD approach as compared to
standard RMF models. Of course, the maximum mass constraint could be used directly
in the determination of the model parameters. But also 
additional constraints at high densities, e.g.\ from heavy-ion collisions,
might help to reduce the uncertainties in the extrapolation from low to high 
densities in the future.

\begin{table}[t]
 \caption{\label{tab:2} Maximum mass, corresponding radius and central
 density of neutrons stars in the DD-NLD models with different parametrizations.}
\vspace{2mm}
 \centering
 \begin{tabular}{lcccc}
 \toprule
 model  &  $M_{\rm max}$ [$M_{\rm sol}$] & $R(M_{\rm max})$ [km] &
 $n_{\rm central}(M_{\rm max})$ [fm${}^{-3}$] \\
 \midrule
 D1  & 2.37 & 11.57 & 0.88  \cr
 D2, $\Lambda = 500$~MeV & 1.48 & 11.75 & 0.95 \\
 D2, $\Lambda = 400$~MeV & 1.36 & 11.63 & 0.97 \\
 D3, $\Lambda = 700$~MeV & 1.26 & 10.13 & 1.41 \\
 D3, $\Lambda = 600$~MeV & 1.19 &  9.99 & 1.46 \\
 \bottomrule
\end{tabular}
\end{table}

\section{Conclusions}
\label{sec:con}

There are several aspects that have to be taken into account in
constraining models of dense matter for the application to neutron
stars. In phenomenological models, the characteristic saturation
properties of nuclear matter and the density dependence of the effective
interaction are usually addressed. However, less attention is paid to
its energy or momentum dependence. 

Introducing non-linear derivative
couplings into RMF models, it is possible to generate an energy
dependence of the nucleon self-energies such that the optical
potential in nuclear matter, which is extracted in Dirac phenomenology
from elastic proton-nucleus scattering experiments, can be well described
up to energies of $1$~GeV.
Considering density-dependent nucleon-meson couplings at the same time,
a very flexible model is obtained. Its parameters can be fitted to the
usual nuclear matter constraints even for different functional forms
of the energy dependent couplings.

In the current version of the DD-NLD model, the
energy dependence of the self-energies causes a softening of the
EoS at high densities, both for symmetric nuclear matter and pure
neutron matter. This effect is independent of the appearance
of additional degrees of freedom, such as hyperons
or deltas. As a result, it becomes more difficult to obtain
very massive neutron stars consistent with the observational
constraints.  In the present work, the density dependence of the 
$\omega$ meson coupling was kept fixed in the parametrizations and only 
a few choices for the energy dependence were tested. Only constraints near
the nuclear saturation density were used to determine the model parameters.
There remains 
substantial freedom in the model to be explored in order to find 
a suitable parametrization that is consistent with the optical potential and
maximum neutron mass constraint.
Nevertheless, the results of the our study indicate that the optical
potential constraint has to be taken seriously into account in the
development of realistic phenomenological models for dense matter.

 RMF models with self-energies that explicitly depend on the 
nucleon momentum or energy can be applied to the simulation of
heavy-ion collisions using relativistic transport approaches. Here the
equation of state can be tested at supra-saturation densities. It is well known
that an energy/momentum dependence of the effective in-medium interaction is mandatory
for a proper description and analysis of experimental data.
Such an approach can help to constrain the parameters of the present model at densities
that are not accessible in the description of finite nuclei.

In our work, an explicit energy dependence of the
nucleon-meson couplings was favored. It allows to apply the DD-NLD
approach to the description of nuclei without major difficulties.
Such an additional investigation of the model will permit a better
control on the parameters.  In particular, the interplay between the
choice of the functional form of the energy dependence, the cutoff parameters and the 
density dependence of the couplings can be studied with 
its impact for the prediction of maximum neutron star masses. 
With a larger number of observables as constraints
it can be expected that firmer conclusions can be drawn on the compatibility of
nuclear matter and finite nuclei descriptions in the DD-NLD model.
Work in this direction is in progress.

\section*{Acknowledgements}
This work was supported by the Helmholtz Association (HGF) through the 
Nuclear Astrophysics Virtual Institute (NAVI, VH-VI-417). S.A. acknowledges
support from the Helmholtz Graduate School for Hadron and Ion Research
(HGS-HIRe for FAIR).

\appendix

\section{Densities and thermodynamic quantities in the DD-NLD model}
\label{sec:app_a}

The proton and neutron densities (\ref{eq:n_i}) in the DD-NLD model
are easily calculated through a momentum integration of the relevant
integral. For other quantities, however, it is more convenient to
introduce an energy integration by substitution because the scalar
($S_{i}$) and vector ($V_{i}$)
potentials are explicit functions of the energy $E_{i}$ of the nucleon
$i$. With the dispersion relation (\ref{eq:disp}) we obtain 
\begin{equation}
 p = \sqrt{\left[ E_{i}-V_{i}(E_{i})\right]^{2} - \left[ m_{i} - S_{i}(E_{i})\right]^{2}}
\end{equation}
and the derivative
\begin{equation}
\label{eq:dpdE}
 \frac{dp}{dE_{i}} = \frac{1}{p} \left[ \left( E_{i} - V_{i} \right) 
 \left( 1 -  \frac{dV_{i}}{dE_{i}}\right)
 + \left(m_{i}-S_{i} \right)
   \frac{dS_{i}}{dE_{i}}  \right] = \frac{\Pi^{0}_{i}}{p}
\end{equation}
with
\begin{equation}
 \frac{dV_{i}}{dE_{i}} = \left( \Gamma_{\omega}\omega
 + \Gamma_{\rho} \tau_{3,i} \rho \right) \frac{dD}{dE_{i}} 
\end{equation}
and
\begin{equation}
 \frac{dS_{i}}{dE_{i}} = \Gamma_{\sigma} \sigma \frac{dD}{dE_{i}} \: .
\end{equation}
Introducing the scalar source densities
\begin{eqnarray}
 n_{i}^{(sD)} & = & \kappa_{i} \int_{0}^{p_{Fi}}
 \frac{d^{3}p}{(2\pi)^{3}} \: \frac{m_{i}^{\ast}}{\Pi_{i}^{0}} D(E_{i})
 \\ \nonumber
 & = & \frac{\kappa_{i}}{2\pi^{2}} \int_{E_{i}^{(\rm
     min)}}^{E_{i}^{(\rm max)}} dE_{i} \: p(E_{i})
 \left[m_{i} - S_{i}(E_{i})\right] D(E_{i})
\end{eqnarray}
and vector source densities
\begin{eqnarray}
 n_{i}^{(vD)} & = & \kappa_{i} \int_{0}^{p_{Fi}}
 \frac{d^{3}p}{(2\pi)^{3}} \:  \frac{E_{i}^{\ast}}{\Pi_{i}^{0}} D(E_{i})
 \\ \nonumber
 & = & \frac{\kappa_{i}}{2\pi^{2}} \int_{E_{i}^{(\rm
     min)}}^{E_{i}^{(\rm max)}} dE_{i} \: p(E_{i})
 \left[E - V_{i}(E_{i})\right] D(E_{i}) \: ,
\end{eqnarray}
the total source densities
\begin{eqnarray}
 n_{\sigma} & = & n_{p}^{(sD)} + n_{n}^{(sD)}
 \\ 
 n_{\omega} & = & n_{p}^{(vD)} + n_{n}^{(vD)}
 \\ 
 n_{\rho} & = & n_{p}^{(vD)} - n_{n}^{(vD)}
\end{eqnarray}
in the field equations (\ref{eq:feq_sigma}), (\ref{eq:feq_omega}), and 
(\ref{eq:feq_rho}) are found. 
The lower and upper boundaries of the integrals are determined by solving
the equations
\begin{equation}
 E_{i}^{(\rm min)}  
 = \left| m_{i} - S_{i}(E_{i}^{(\rm min)}) \right|
 +  V_{i}(E_{i}^{(\rm min)})
\end{equation}
and
\begin{equation}
 E_{i}^{(\rm max)}  
 = \sqrt{ p_{Fi}^{2} + \left[ m_{i} - S_{i}(E_{i}^{(\rm max)}) \right]^{2}}
 +  V_{i}(E_{i}^{(\rm max)}) \: ,
\end{equation}
respectively, with the Fermi momenta $p_{Fi}$ from equation (\ref{eq:n_i}).
The argument of the coupling functions
$n_{v}=n_{p}^{(v)}+n_{n}^{(v)}$ can be obtained from
\begin{eqnarray}
 n_{i}^{(v)} & = & \kappa_{i} \int_{0}^{p_{Fi}}
 \frac{d^{3}p}{(2\pi)^{3}} \:  \frac{E_{i}^{\ast}}{\Pi_{i}^{0}} 
 \\ \nonumber 
 & = & \frac{\kappa_{i}}{2\pi^{2}} \int_{E_{i}^{(\rm
     min)}}^{E_{i}^{(\rm max)}} dE_{i} \: 
 p(E_{i})  \left[E_{i} - V_{i}(E_{i})\right] \: .
\end{eqnarray}
The energy density assumes the form
\begin{eqnarray}
 \varepsilon & = & \sum_{i=p,n} \kappa_{i} \int_{0}^{p_{Fi}}
 \frac{d^{3}p}{(2\pi)^{3}} \:  E_{i} - \langle \mathcal{L} \rangle
 \\ \nonumber & = &
 \frac{\kappa_{i}}{2\pi^{2}} \int_{E_{i}^{(\rm
     min)}}^{E_{i}^{(\rm max)}} dE_{i} \: 
 p(E_{i}) 
 \Pi_{i}^{0}(E_{i})  E_{i} - \langle \mathcal{L} \rangle
\end{eqnarray}
and the pressure is given by
\begin{eqnarray}
 p & = & \frac{1}{3} \sum_{i=p,n} \kappa_{i} \int_{0}^{p_{Fi}}
 \frac{d^{3}p}{(2\pi)^{3}} \:  \frac{p^{2}}{\Pi_{i}^{0}}
 + \langle \mathcal{L} \rangle
 \\ \nonumber & = &
 \sum_{i=p,n} \frac{\kappa_{i}}{6\pi^{2}} \int_{E_{i}^{(\rm
     min)}}^{E_{i}^{(\rm max)}} dE_{i} \: \left[p(E_{i})\right]^{3}
 + \langle \mathcal{L} \rangle
\end{eqnarray}
with
\begin{equation}
 \langle \mathcal{L} \rangle = 
 \frac{1}{2} \left( \Gamma_{\omega} \omega n_{\omega}
  + \Gamma_{\rho} \rho n_{\rho}
 - \Gamma_{\sigma} \sigma n_{\sigma}  \right)
 +  \left( \Gamma_{\omega}^{\prime} \omega n_{\omega}
 + \Gamma_{\rho}^{\prime} \rho n_{\rho}
 - \Gamma_{\sigma}^{\prime} \sigma n_{\sigma}  \right) n_{v} \: .
\end{equation}
Using a partial integration and equation (\ref{eq:dpdE}), the thermodynamic identity
\begin{eqnarray}
 \varepsilon + p & = & 
 \sum_{i=p,n} \frac{\kappa_{i}}{(2\pi)^{3}} 
 \left. \frac{p^{3}}{3} E_{i} \right|_{0}^{p_{Fi}}
 \\ \nonumber & & 
 + \sum_{i=p,n} \frac{\kappa_{i}}{(2\pi)^{3}}
 \left\{
 - \frac{1}{3} \int_{0}^{p_{Fi}} d^{3}p \:  p \frac{dE_{i}}{dp}  
 + \frac{1}{3} \sum_{i=p,n} \kappa_{i} \int_{0}^{p_{Fi}}
 d^{3}p \:  \frac{p^{2}}{\Pi_{i}^{0}} \right\}
 \\ \nonumber & = & 
 \sum_{i=p,n} \mu_{i} n_{i}
\end{eqnarray}
with the chemical potential $\mu_{i} = E_{i}(p_{Fi})$ is easily confirmed.

\section*{References}

\bibliography{paper}{}
\end{document}